\documentclass[aps,prl,twocolumn,superscriptaddress]{revtex4}
\usepackage[T1]{fontenc}
\usepackage{amsmath} 
\usepackage{mathtools}
\usepackage{physics} 
\usepackage{amsfonts} 
\usepackage{booktabs} 
\usepackage{graphicx} 
\usepackage{float} 
\usepackage{multirow}
\usepackage[caption=false]{subfig}
\usepackage{hyperref}
\usepackage{xcolor}
\usepackage{upgreek}
\begin{document}
\title{Experimental realization of local-to-global noise transition in a two-qubit optical simulator}
\author{Claudia Benedetti, Valeria Vento, Stefano Olivares, Matteo G. A. Paris, Simone Cialdi}
\affiliation{Dipartimento di Fisica "Aldo Pontremoli", 
Universit\`a degli Studi di Milano, 
I-20133 Milano, Italia}
\begin{abstract}
We demonstrate the transition from local to global noise 
in a two-qubit all-optical quantum simulator subject to 
classical random fluctuations. Qubits are encoded in the 
polarization degree of freedom of two entangled photons 
generated by parametric down-conversion (PDC) while the environment 
is implemented using their spatial degrees of freedom.
The ability to manipulate with high accuracy the  number of 
correlated pixels of a spatial-light-modulator 
and the spectral PDC width, allows us to control the transition from 
a scenario where the qubits are
embedded in local environments to the situation where they 
are subject to the same global noise.
We witness the transition by monitoring the decoherence
of the two-qubit state.
\end{abstract}
\maketitle
Quantum simulators are controllable quantum systems, usually made of qubits, 
able to mimic the dynamics of other, less controllable, quantum systems 
\cite{nori09, nori14}. Quantum simulators make it possible to 
design and control the dynamics of complex systems with a large 
number of degrees of freedom, or with stochastic components
\cite{jaksch05,kassal11,Walther12,blatt12,houck12}. In turn, 
open quantum systems represent a fundamental testbed to 
assess the reliability and the power of a quantum simulator. 
The external environment may be  described 
either as a quantum bath, or a classical random field which, in generale, 
lead to different system evolutions. However, in the case of pure 
dephasing, the effects of a quantum bath are equivalent to those 
provoked by random  fluctuations \cite{crow14}. 
For this reason, together with the fact that it is an ubiquitous 
source of decoherence that jeopardizes quantum features, dephasing  
noise plays a prominent role in the study of open quantum systems.
\par
Pioneering works on the controlled simulation of  single-qubit 
dephasing channels appeared few years ago  \cite{cialdi17, liu18}, 
whereas the realisation of multi-qubit simulators is still missing. 
In fact, the  simulation of multi-qubit systems is  not a mere 
extension of the single-qubit case since composite systems present 
features that are absent in the single-component case, e.g. 
entanglement \cite{yu03, benedetti12,benedetti13,costa16,daniotti18}. 
Moreover, multipartite systems allow us to analyze the effects 
of a global source of noise against those due to 
local environments. Understanding the properties of the 
local-to-global (LtG) noise transition is in turn 
a key task in quantum information, both for quantum and 
classical environments, since it sheds light on the 
mechanisms governing the interaction between the quantum 
system and its environment, providing tools to control 
decoherence 
\cite{fisher09,bruss11,lofranco13,addis13,rossi14,zambrini17}.
\par
We present here an all-optical implementation of the whole
class of two-qubit dephasing channels arising from the interaction with 
a classically fluctuating environment. The qubits are encoded in the polarization degree of freedom of a photon-pair generated by 
parameteric-down-conversion (PDC), while the 
spatial degrees of freedom are used to implement the environment. 
Different realizations of the noise are randomly generated and 
imprinted on the qubits through a spatial-light modulator
(SLM). The ensemble average is then performed by collecting the 
photons with a multimode fiber. With our simulator there is no need 
to work at cryogenic temperatures and we are able to simulate 
{\em any} conceivable form of the environmental noise, independently 
on its spectrum. 
\par
\begin{figure}[!h]
\includegraphics[width=0.83\columnwidth]{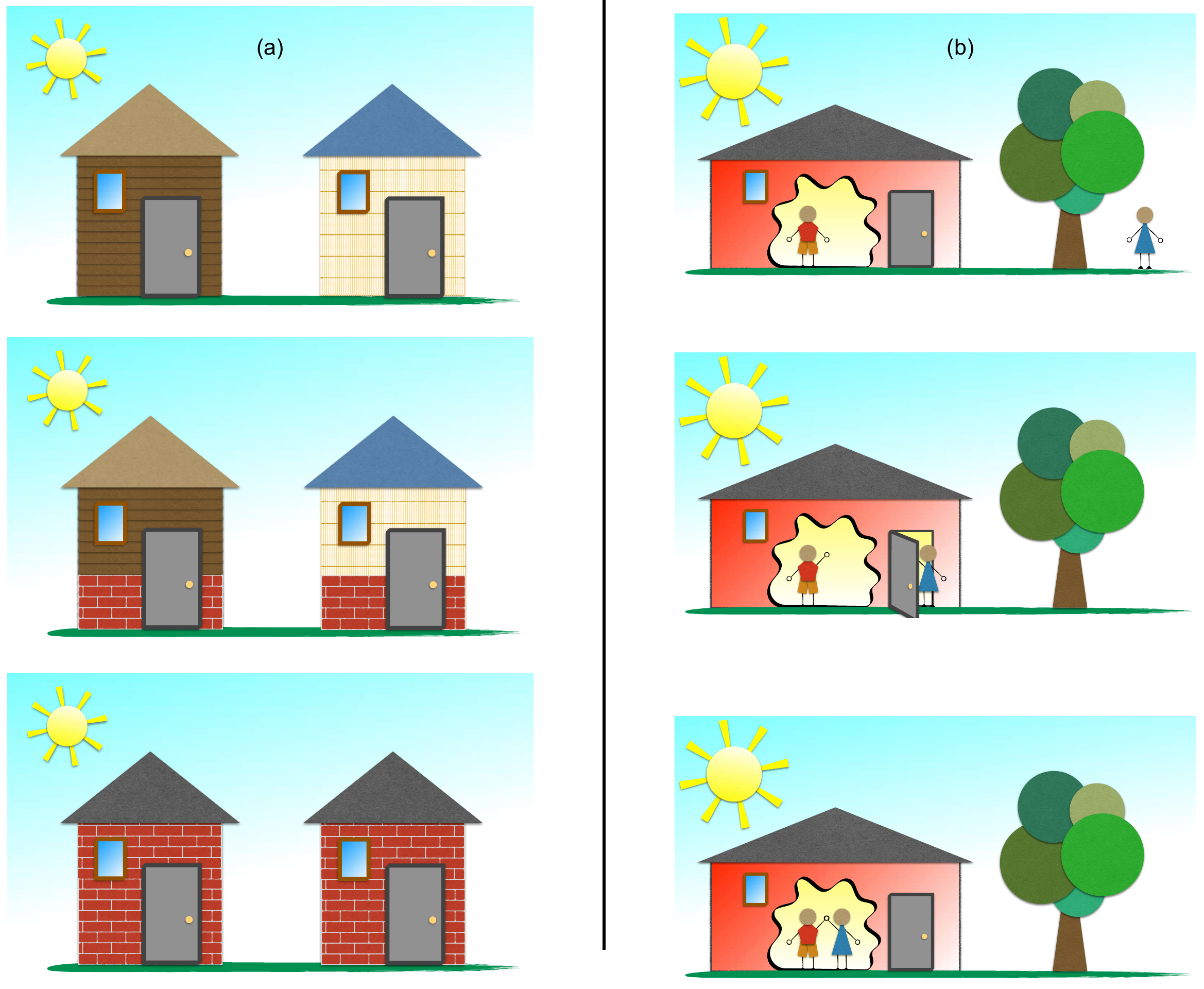}
\caption{Cartoon description of the two scenarios leading to 
the LtG transition. On the left: LtG transition after establishing 
correlations between the two environments. On the right: LtG transition
as the qubits approach each other. \label{f:pict}}
\end{figure}

In particular, here we exploit our simulator to demonstrate 
the transition from a local-environment scenario, where each 
qubit is subject to an independent source of noise, to a global 
environment where both qubits feel the same synchronous random 
fluctuations. There are two different mechanisms that may lead 
to this transition. The first one appears when two local
environments become correlated due to the action of some external
agent, and one moves from local to global noise as the two 
environments become fully correlated. In the second scenario, 
the two qubits are placed in the same environment, but at a distance 
that is much larger than the correlation length of the noise.  As
the distance between the qubits is reduced, they start to feel
similar environments, until they are within the correlation length 
of the environment and thus subject to the same common source of noise. 
The two situations are illustrated in Fig. \ref{f:pict}. In the first 
case, two initially different environments (cottages) become 
gradually identical as far as correlations are established 
(by putting bricks), whereas in the second case 
the two qubits (persons) are initially far apart, but they end up 
feeling the same environment (cottage) as long as their distance 
is reduced. 
\par
The dephasing map of two  non-interacting qubits arising
from a classical environment is generated by the dimensionless 
Hamiltonian:
\begin{equation}
 H(t)=X_1(t)\,\sigma_z^{(1)}\otimes \mathbb{I}^{(2)}+X_2(t)\,
 \mathbb{I}^{(1)}\otimes \sigma_z^{(2)},
 \label{ham}
\end{equation}
where $\sigma_z$ is the Pauli matrix, $\mathbb{I}$ the identity matrix,  $X(t)$
 is a stochastic process and the labels 1 and 2 denote the two qubits.
 Since our aim is to give a proof-of-principle of the LtG transition and we are not
 interested at this stage in the specific form of the noise,
 we fix the stochastic process to be a
 random telegraph noise (RTN). 
 It follows that $X_k(t)=\pm 1$ is a \textcolor{black}{dichotomous} variable which
 jumps between two values with a certain switching rate $\gamma$
that determines the correlation length of the noise
through the
 autocorrelation function $C(t)=\langle X(t)X(0)\rangle=e^{-2\gamma t}$.
 The symbol $\langle \dots\rangle$ denotes the ensemble average over all
 possible realizations of the  RTN.
 In Eq. \eqref{ham} it is possible to identify two complemtary regimes:
 If $X_1(t)$ and $X_2(t)$ are two identical but independent processes,
 then we are in the presence of local environments and each qubit is 
 subject to its own noise. On the other hand, if $X_1(t)=X_2(t)$, at all times,
 then the fluctuations are synchronized (perfectly correlated) and the qubits  
 interact with the same global environment. 
 The generated dynamics in 
 these two scenarios are very different and this can be 
 witnessed, for example, by looking at the behavior of entanglement. 
What happens in between these regimes is unexplored territory.
\par {\it Theoretical model} $-$
In order to address the local-to-global noise transition, we first 
need to compute the two qubits dynamics in the presence of
classical noise. 
 Starting from an initial Bell state 
 $\ket{\psi_0}=\frac{1}{\sqrt{2}}(\ket{\text{\small HH}}+\ket{\text{\small VV}})$, 
 the system  density matrix is obtained as
 $\rho(t)=\langle U(t)\rho_0U^{\dagger}(t)\rangle$, where 
  $\rho_0=\ketbra{\psi_0}{\psi_0}$, 
 $U(t)=\exp\left[-i\int\! H(s)ds\right]=\exp[-i\varphi(t)]$ , $\varphi(t)=\int H(s)ds$ 
is  the noise phase.
 The elements of $\rho(t)$ in the polarization basis
 $\{ \ket{\text{\small HH}},\ket{\text{\small HV}},\ket{\text{\small VH}},\ket{\text{\small VV}}\}$ are:
$
\rho_{\text{\tiny 11}}(t)=\rho_{\text{\tiny 44}}(t)=\frac12$ and $
\rho_{\text{\tiny 14}}(t)=\rho^*_{\text{\tiny 41}}(t)=\frac12 \Gamma(t)$,
and all other elements are zero.
The coherence factor
$\Gamma(t)$ depends on the nature and the correlations of the noises.
In particular,  it was shown that for local (LE) and global environments (GE) 
the coherence factor takes the forms respectively:
\begin{align}
\Gamma_{ \text{\tiny LE}}(t)=\langle e^{2i\varphi(t)}\rangle^2
\quad
\Gamma_{ \text{\tiny GE}}(t)=\langle e^{4i\varphi(t)}\rangle
\label{gamma}
\end{align}
where the averages of the exponential moments are given by $\langle e^{m \,i\varphi(t)}\rangle=e^{-\gamma t}(\cosh\delta_m t + \gamma/\delta_m \sinh \delta_m t)$ with $\delta_m=\sqrt{\gamma^2-m^2}$.
The entanglement $\mathcal{E}$ between the qubits is given by $\mathcal{E} (
t)=|\Gamma (t)|$ in both LE and GE cases.
\par
The realization of the qubit state $\rho(t)$ requires the 
simultaneous generation of a large number of stochastic trajectories of the noise.
Our experimental apparatus allows us to obtain the average over the realizations
in parallel, exploiting the \textcolor{black}{spatial} and the 
\textcolor{black}{spectral} degrees of freedom of the
photons. 
In particular, in our experimental setup the following state is generated 
\begin{align}
\ket{\psi_{\text{\tiny SE}}(t)}=\textcolor{black}{\frac{1}{\sqrt{2}}}  \int\! &dx_1dx_2 f(x_1,x_2)\,
\Big[ \ket{\text{\small H}\,x_1\text{\small H}\,x_2}\nonumber\\
&+\,e^{i [\varphi_1(x_1,t)+\varphi_2(x_2,t)]}\ket{\text{\small V}\,x_1\text{\small V}\,x_2} \Big]
\label{state1}
\end{align}
where $f(x_1,x_2)$ is the \textcolor{black}{spatial} correlation function between 
the two photons, the $\textcolor{black}{\varphi_k(x_k,t)}$'s are  the noise phases, and
$\ket{\text{\small P}_1x_1\text{\small P}_2x_2}$ denotes a 
state where the photon 1(2) has polarization P$_1$ (P$_2$) and is in position $x_1$ ($x_2$). 
\textcolor{black}{In this scenario, 
the stochastic trajectories are encoded in the spatial degree of freedom and 
the state $\rho (t)$ is obtained by tracing out $x_1$ and $x_2$. 
Finally, as we will show below, we employ the spectral degree of freedom
to define the degree of spatial correlation between the two photons}. 
\begin{figure}[t!]
\centering
\includegraphics[width=0.8\columnwidth]{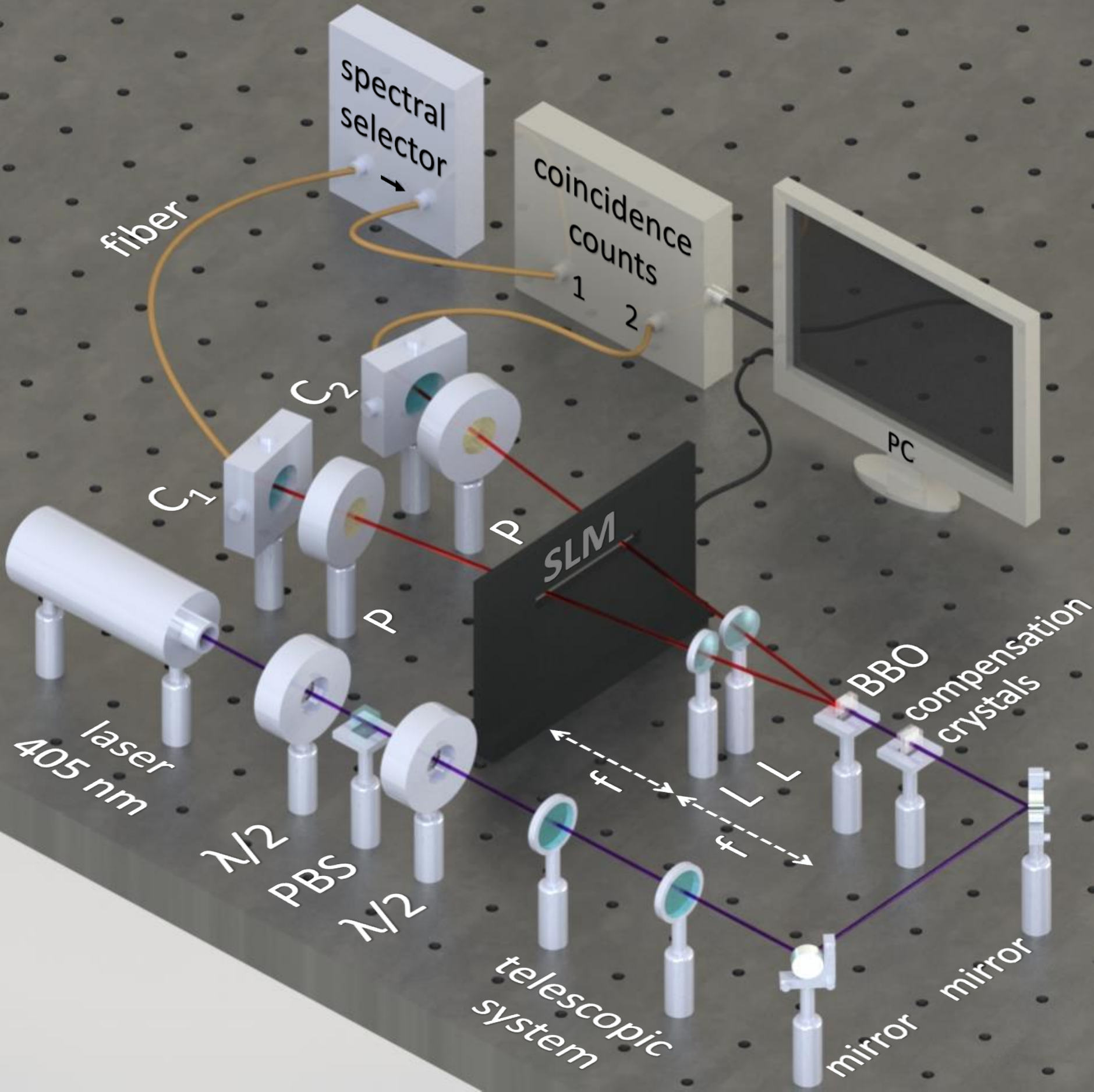}
\caption{
Schematic diagram of the experimental setup. A $405$-nm cw laser diode generates a pump beam which passes through a half-wave plate ($\lambda/2$), a polarizing beam-splitter cube (PBS), and another half-wave plate. Then it is collimated by a telescopic system composed by two lenses $L_{t0}$ and $L_{t1}$.
 The beam passes through a series of compensation crystals and then it interacts with two 1-mm long BBO crystals generating photons centered at $810$ nm via PDC. Each branch passes through a lenses $L$ with focal $f$, the spatial light modulator (SLM) and a polarizer (P). Photons are finally focused into two multimode fibers through the couplers $C_1$ and $C_2$: the first is directly linked to an homemade single-photon counting module, the second is sent to a spectral selector and then to the counting module}
\label{fig:setup}
\end{figure}
\par
{\it Experimental realization} $-$
Our experimental setup is schematically depicted 
in Fig. \!\ref{fig:setup}. The pump is generated from a 405-nm cw 
InGaN laser diode. The laser beam passes through an amplitude modulator, 
composed by a half-wave plate and a polarizing beam-splitter (PBS), and then through another 
half-wave plate to set the polarization. Subsequently, the polarized 405-nm beam goes through a telescopic
system composed by two lenses $L_{t0}$ and $L_{t1}$ with respective focal lengths
$f_{t0}=100$ mm and $f_{t1}=75$ mm, with the double purpose of collimating the beam and optimizing its 
size \textcolor{black}{in order to maximize the detection efficiency}. 
After the telescopic system there are three 1-mm long crystals that 
compensate the  delay time introduced by the PDC crystals. 
\textcolor{black}{In order to generate the entangled state in the 
polarization we use two $1$ mm long crystals of  beta-barium borate 
(BBO) \cite{kwiat99}. On each branch of the PDC}, a lens $L$ with 
focal length $f=200$ mm at 810 nm 
is placed at a distance $f$ from the BBO. On the Fourier's plane (\textcolor{black}{at distance $f$ from the lens $L$}) there is the SLM which is a 1D liquid crystal mask with 640 pixels of width 100 $\upmu$m/pixel. Each pixel imprint a computer-generated 
phase on the \textcolor{black}{horizontal}-polarized component. Photons then pass through polarizers and are 
finally focused into a multimode fiber through the couplers $C_1$ and $C_2$ and into a photon-counting
module. The signal beam, before the detection, passes through a spectral selector, which consists of two gratings and two lenses building 
an optical 4f system. In the Fourier plane of this
apparatus, we use a mechanical slit in order to select the 
desired spectral width \cite{anelli}. 
\par
Since each pixel of the SLM has a finite width, we may substitute the
integral with a sum over the pixels positions in  Eq. \eqref{state1}. 
Then, by taking the partial trace over the spatial degrees of freedom, 
we obtain $
\rho_{\text{\tiny S}}(t)  =  \frac12 ( \ketbra{\text{\small HH}} + \ketbra{\text{\small VV}}  + p\,\Gamma(t) \ket{\text{\small HH}}\!\!\bra{\text{\small VV}}+ p\,\Gamma^*(t) \ket{\text{\small VV}}\!\!\bra{\text{\small HH}})$
\textcolor{black}{where $p$ is a parameter quantifying the entanglement in the initial 
state $\rho_{\text{\tiny S}}(0)=p\,\rho_0 + (1-p) \rho_{\text{\tiny mix}}$,
where $\rho_{\text{\tiny mix}}=\frac{1}{2} \left( \ketbra{\text{\small HH}} + \ketbra{\text{\small VV}} \right)$. In our case $p$ is close to 1 \cite{supp}
and the procedure we use to purify the state is described in \cite{cialdi10, cialdi10B}}. 
The decoherence function $\Gamma(t)$ depends on 
the spatial correlations between the two photons and on the stochastic realizations:
\begin{equation}
\Gamma(t)=\sum_{jk}|f_{jk}|^2 e^{i [\varphi_1(x_{1j},\textcolor{black}{t})+\varphi_2(x_{2k},t)]}.
\end{equation}
where the distribution
\begin{align}
|f_{jk}|^2 = &N_0 \exp\left\{
-\frac{2[(j-j_0)-(k-k_0)]^n}{w^n_{\text{\tiny cp}}}\right\}\nonumber\\
&\times\exp\left\{-\frac{2(j-j_0)^2}{w_p^2}-\frac{2(k-k_0)^2}{w^2_p}\right\}
\label{ff}
\end{align}
takes into account the size of the coupled PDC $w_p$ and spatial correlation 
between the photons (i.e. the number of correlated pixels) $w_{\text{\tiny cp}}$  
(See the supplementary material for details and the detivation).
The first factor is a super-Gaussian of order $n$, while $j_0$ and $k_0$ are 
the central pixes on the SLM for each PDC branch.
Finally $N_0$ is a normalization factor in order to assure that $\sum_{jk}|f_{jk}|^2=1$.
It is now clear that we may simulate the LtG transition using two different 
strategies, either by controlling the realizations of the noise on the two paths 
of the PDC, or by tuning the number of correlated pixels.
\par
{\it Results} $-$ 
Let us start by explaining 
the role of the SLM in encoding the stochastic process into the pixels.
We figuratively divide the SLM in two parts, both made of 320 pixels. 
The first set is dedicated to the first qubit and the pixels are indexed by an integer $j$
that goes from $0$ to $319$. 
The second part is dedicated to the second qubit and the 
pixels are labeled by $k$ that goes from $320$ to $639$.
Called $d$ the width of the pixel, the two positions $x_1$ and $x_2$ are: $x_{1j}=d \,j$, $x_{2k}=d(640-k)$. 
This allows us to directly consider  the 
simmetry of the spatial correlations between the photons in the notation.

\textcolor{black}{The first step in order to send the same noise on the correlated pixels in the two parts of the mask is to experimentally find out  the central pixels $j_0$ and $k_0$.
The central pixels are the reference for the definition of the phases $\varphi_1$ and $\varphi_2$. In particular we set:
\begin{align}
 \varphi_1(x_{1, j_0+\Delta},t) = \varphi_2(x_{2, k_0+\Delta},t) = \varphi(\Delta,t)
\label{phases}
\end{align}
where $\Delta$ is an integer shift with respect the references and $\varphi$ is a phase 
function \textcolor{black}{defined over $320$ points}.}
The first method we use to simulate te LtG transition consists in introducing
an integer shift $\delta$ on the array of phases $\varphi_2$ imprinted on the second side of the SLM. 
Setting $\delta=0$ the correlated pixels see the same noise and we mimic
 the case where the environments are fully correlated.
When  $\delta$ is increased, the two environments become 
progressively less correlated. It follows, that as the value of $\delta$ is
decreased from a large value to zero, we obtain the LtG transition.
The function ${\Gamma}$ is now a function of $\delta$ and we have:
\begin{align}
\Gamma(\delta,t)=\sum_{jk}|f_{jk}|^2 e^{i [\varphi_1(x_{1j},t)+\varphi_2(x_{2k+\delta},t)]}\,.
\end{align}
Since this technique is effective for
spatial correlation lengths $w_{\text{\tiny cp}}$ smaller or equal
with respect to the typical spatial variation of the function $\varphi$,
we use a spectral width
of $\Delta \lambda=15$ nm, resulting in a $w_{\text{\tiny cp}}$ of about $3$ pixels \cite{supp} and use a function $\varphi$ that changes value every $3$ pixels.
By reducing the spectrum width it is possible to obtain smaller values of 
$w_{\text{\tiny cp}}$ at the price of reducing counts and, in turn, increasing fluctuations.
\begin{figure}[h!]
\centering
\includegraphics[width=0.98\columnwidth]{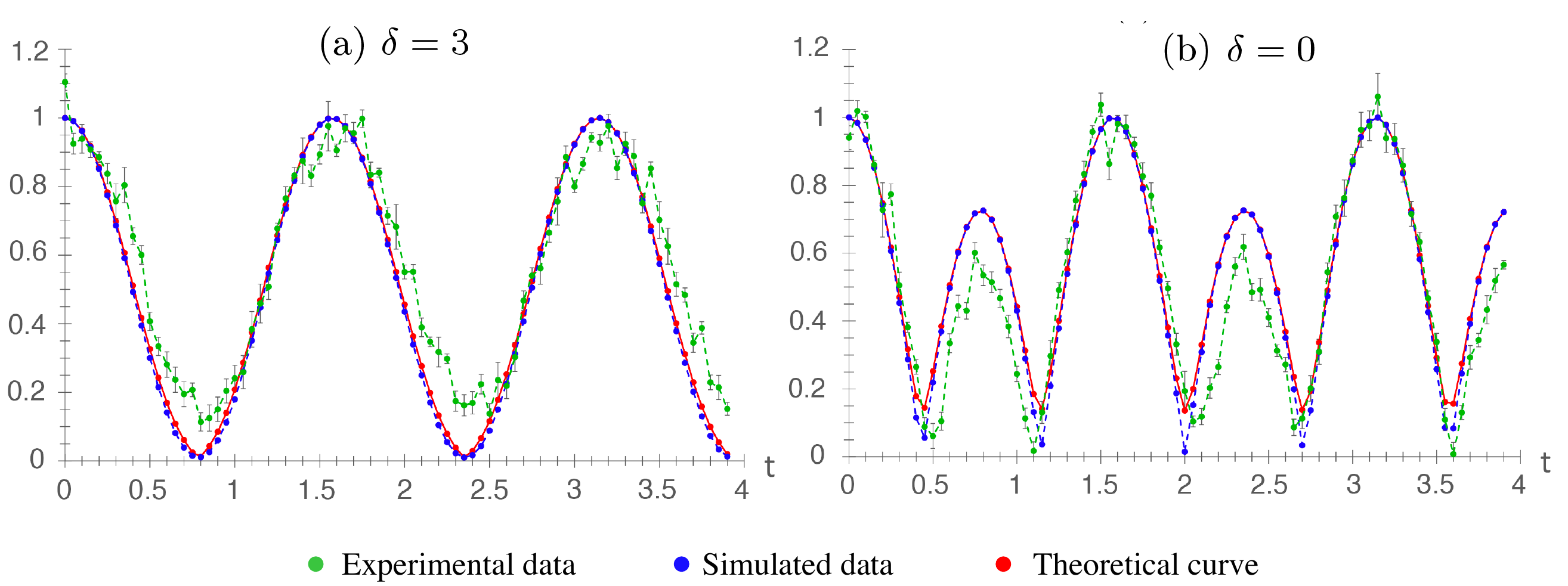}
\caption{Local-to-global noise transition by translation 
of the realizations $\varphi(t)$
for the RTN with $\gamma=0$ and $\Delta\lambda=15$ nm.  
The green curves represent the experimental data, while the blue curves are the corresponding simulations for $|\Re[{\Gamma(t)}]|$. 
The red curve is $|\Gamma(t)|$. In the left panel we show results for 
$\delta=3$, i.e. the two environments are not correlated. The right panel 
is $\delta=0$, i.e. fully correlated environments.}
\label{exp1}
\end{figure}
\par
At first, let us consider the experimental realization of the 
local-to-global transition in the case of the RTN with $\gamma=0$ 
using the $\delta$ shift-technique.  Upon imposing $\delta=0$ we obtain two fully correlated 
environments, and using Eq.  \eqref{gamma} we have 
$|\Gamma_{ \text{\tiny GE}}(t)|=
|\langle e^{4i\varphi(t)}\rangle| = |\cos(4t)|$.
Indeed, due to the RTN noise, the only two possible values for $\varphi$ 
are $t$ and $-t$. 
The blue and the red curves in Figure \ref{exp1}b.
are respectively $|{\Gamma}|$ and 
$|\Re{{\Gamma}}|$ obtained theoretically using  the experimental values for  
$w_{\text{\tiny cp}}$ and $w_{p}$  $3$ and $20$ pixels respectively.
The comparison between the red and the blue curves shows that the $\Gamma$ function is real, as it would be in the ideal case with 
an infinite number of realizations. In order to obtain this result it is necessary to select the $\varphi$ function 
with a balanced number of positive and negative realization at time $t=0$.
When $\delta=3$ (Figure \ref{exp1}a)  the environments are not correlated.
In this case the two qubits see two different phases (indeed the shift $\delta$ makes the two quantities $\varphi_1$ and $\varphi_2$
completely different at the correlated positions). Using Eq. \eqref{gamma} we
have $|\Gamma_{ \text{\tiny LE}}(t)|=\langle e^{2i\varphi(t)}\rangle^2 = \cos(2t)^2$. Notice that the second peak in  Figure \ref{exp1}b 
does not reach the value 1 due to the undersampling of the 
noise realizations \cite{rossi17}. 
\par
\begin{figure}[!t]
\centering
\includegraphics[width=0.99\columnwidth]{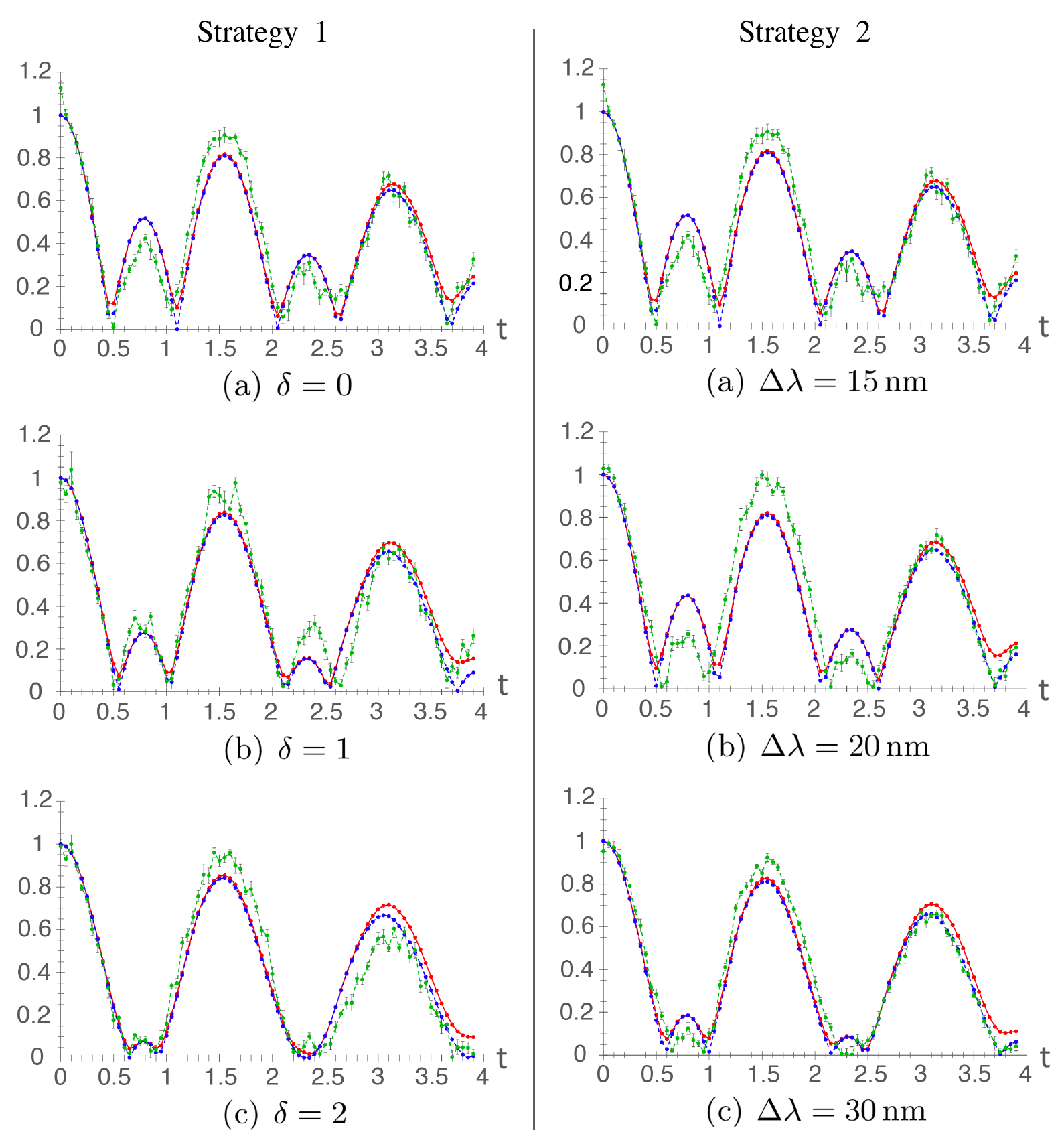}
\caption{Experimental implementation of the transition from global to local noise in the case of RTN with $\gamma=0.12$. In the left column the $\delta$ shift-strategy is used. On the right column the transition is obtained by changing the spectral width $\Delta\lambda$.}
\label{exp2}
\end{figure}

The second strategy to obtain the LtG transition consists in 
increasing the number of correlated pixels $w_{\text{\tiny cp}}$ while fixing 
$\delta=0$ and the number of repeated pixels in the function $\varphi$. In order 
to increase $w_{\text{\tiny cp}}$
we increase the width of the PDC spectrum by acting on the spectral selector 
\cite{supp}.
When the spectral width is $\Delta \lambda =15$ nm, the number of correlated
pixels  is equal to the number of repeated pixel in the $\varphi$ function 
and the two quibit see the same environment.
By progressively increasing the value of $\Delta \lambda$,  $w_{\text{\tiny cp}}$ 
becomes larger than the number of repeated pixels in $\varphi$ and 
the two qubits see different environments. Indeed, on the two photons  
are imprinted different phases.
\\
In Fig. \ref{exp2}, we show the experimental realization 
of the transition, using both techniques.
In the left column the $\delta$-shift method is used, 
on the right column the transition is obtained changing the PDC spectral width.
We note that moving from local noises to a global noise the rising of the new peaks are evident.
Here $\gamma=0.12$ in order to show a case with a non stationary RTN noise.
Fluctuations in the experimental data are
mostly due to the strong dependence on the central pixel. Indeed, 
the center of the PDC beam may be shifted from the center of the 
pixel itself for a fraction of the pixel's length: this has been 
taken into account in the simulations but it changes from one 
measure to another and it not possible to estimate it with 
a sufficient precision.
\par
{\it Conclusions} $-$  We have experimentally demonstrated the 
transition from local-to-global decoherence in an all-optical 
two-qubit quantum simulator subject to classical noise. We 
exploited the spatial degrees of freedom of the PDC photons 
to implement the noise realizations while the photons 
polarizations encoded the two qubits.  In particular, thanks
to the high control of the PDC width and of the spatial 
correlations among pixels of the SLM, we have been able 
to implement two different strategies  for the noise transition,
either involving the building of correlations between environments
or the tuning of the PDC spectral width.
Besides RTN, that we used as a testbed for our simulator, 
any kind of classical noise may be implemented, making our scheme
suitable to simulate a wide range of dynamics involving super- and 
semi-conducting qubits that are of the utmost importance for   quantum technologies.
\par
Our results also paves the way to the realization of
many-qubit simulators, and open up to the chance of 
explore the dynamics of multi-partite entanglement 
as well as to study the robustness of quantum features 
against decoherence.  In multi-qubit systems the LtG transition takes a broader 
meaning with sub-groups of qubits 
that may feel the same noise while others are subject local fluctuations. 
In this case spatial correlations becomes the key element that 
governs the dynamics \cite{borrelli}. More generally, 
the ability to monitor and control the LtG transition is a 
fundamental step in understanding decoherence,
especially in the context of reservoir engineering, where the noise 
is tailored or to improve performances of specific 
protocols \cite{poyatos96,defalco13,manis13}.

$ $ \vfill $ $ \newpage
$ $ \vfill $ $ \newpage
\begin{widetext}
\section*{Supplementary material for "Experimental realization of local-to-global noise transition in a two-qubit optical simulator" by C. Benedetti et al}
\renewcommand{\theequation}{S.\arabic{equation}}
\renewcommand{\thefigure}{S.\arabic{figure}}
\setcounter{equation}{0}
\setcounter{figure}{0}
{\bf Two-photon spatial correlations} $-$
Figure \ref{pdcs}  shows the geometrical configuration of the two photons generated by PDC. The two angles
$\theta_1$ and $\theta_2$ are the angular shifts with respect the PDC central angle $\theta_0$ defined by the phase-matching condition. 
The coordinates
$x_{1(2)}$ and $x_{01(02)}$ are respectively the positions and the references on the SLM plane.
The arrows represent the orientations we use for the axes.
The two-photon state can be written \cite{anelli}:
\begin{eqnarray}
\label{eq:totall}
 |\psi\rangle =  \frac{1}{\sqrt{2}} \int\!  d\omega \,d \theta_1 d \theta_2 
\tilde{A}(\Delta k_{\perp})\, {\mbox{Sinc}}( \Delta k_{\parallel} L/2) 
\left[\ket{{\small{H,\theta_1, \omega}}}\ket{{\small{H,\theta_2,-\omega}}}+e^{i \Phi(\theta_1, \theta_2)}
\ket{{\small{V,\theta_1, \omega}}}\ket{{\small{V,\theta_2,-\omega}}}\right], 
\end{eqnarray}
where, up to first order in frequency and angle, $\Delta k_{\parallel} =  - \omega_p^0 \theta^0 (\theta_1+\theta_2) /(2 c)$
and $\Delta k_{\perp} =  \omega_p^0 (\theta_1-\theta_2) /(2 c) + 2 \theta^0 \omega/c$.
 $\omega_p^0$ is the pump frequency ($405$ nm) and $\omega$ is the frequency shift with respect the PDC central frequency
$\omega_p/2$ ($810$ nm), c is the speed of light and the phase term $\Phi(\theta_1,\theta_2)$ is due to the different optical paths followed by the pairs of photons generated in the first and in the second crystal. 
\begin{figure}[h]
\centering
\includegraphics[width=0.4\textwidth]{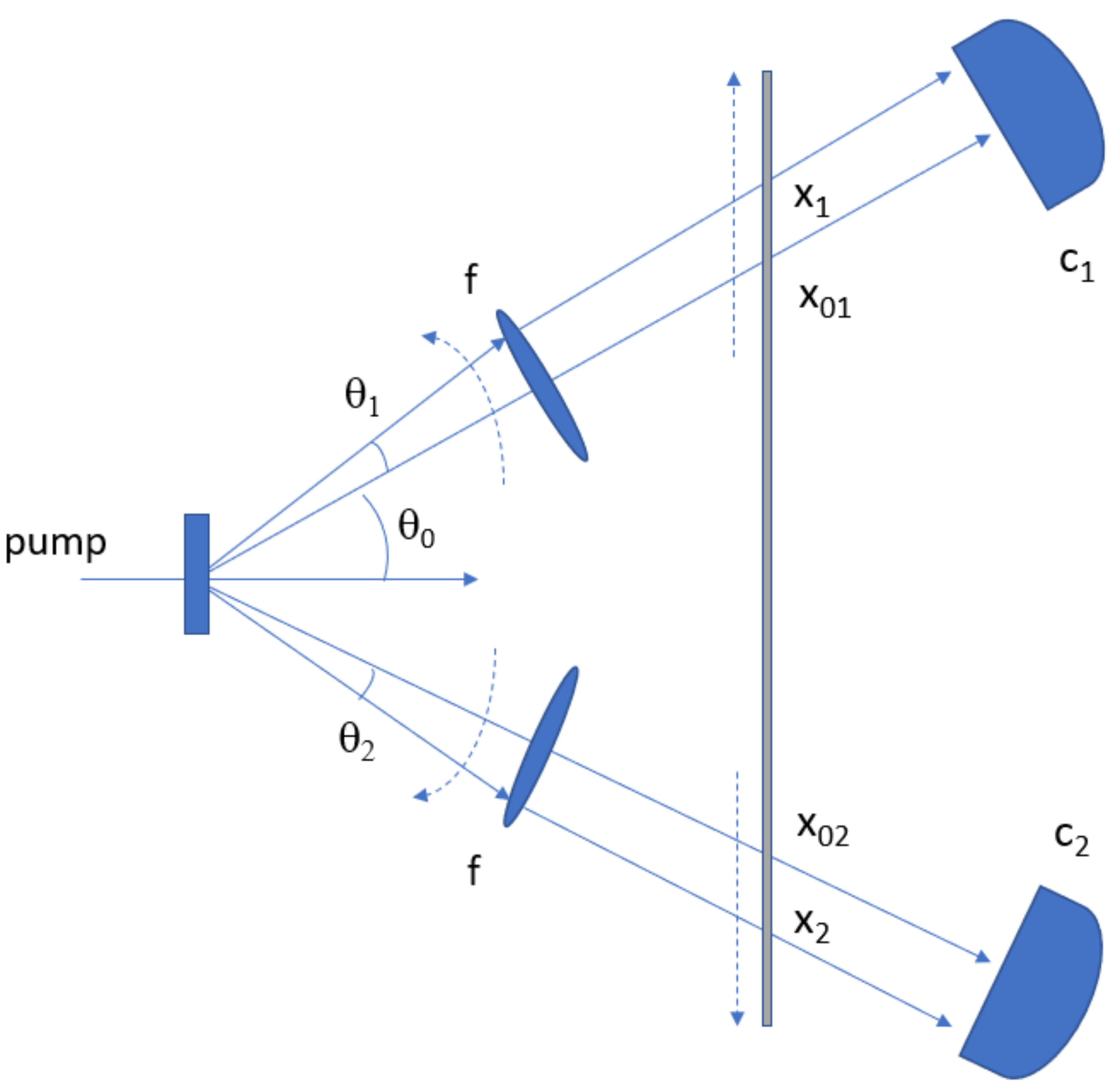}
\caption{The geometry of the parametric downconversion scheme}
\label{pdcs}
\end{figure}
\\
The function $\tilde{A}$ comes from the integration along the transverse coordinate \cite{cialdi12} and it is the Fourier transform of 
the pump spatial amplitude, the Sinc function comes from the integration along the longitudinal coordinate inside the crystal ($L$ is the 
crystal length). $\Delta k_{\parallel}$ and $\Delta k_{\perp}$ are the shifts with respect the phase-matching condition of the longitudinal and the transverse
momentum of the two photons.
Finally, we obtain $\Phi=0$ by using the purification method explained in \cite{cialdi10, cialdi10B}. 
\textcolor{black}{Due to the pump spot dimension (\textcolor{black}{$w_{\text{pump}}\approx 0.6$} mm) and the crystal length ($1$ mm) employed in our experimental configuration, the angular correlations depend mainly on
 the function  $\tilde{A}$ and in turn by $\Delta k_{\perp}$.
  Considering that $\Delta k_{\perp}$ is a function of $\omega$,
it is simple to get that by enlarging the PDC spectrum width we progressively lose the angular correlations. 
Moreover, when the spectrum width goes to $0$,
the angular correlations depend directly on the width of the pump spot
 via its Fourier transform $\tilde{A}$.
In order to obtain both the PDC width $w_p$ and the spatial correlations, we define:
$F(\theta_1, \theta_2) = \int \!d\omega\, |\tilde{A}(\Delta k_{\perp}) {\mbox{Sinc}}( \Delta k_{\parallel} L/2)|^2 $.
Before proceeding with the calculation, we have to switch from angular to spatial coordinates.
Due to the fact that the lens $f$ is placed at distance $f$ both from the crystals and  the SLM, 
we have: $\Delta x_{1,2}=
x_{1,2}-x_{01,02}=f \theta_{1,2}$ and  we can write $F$ as a function of $\Delta x_{1,2}$.
\\ \\
To estimate the PDC width $w_p$ we define the function: \textcolor{black}{$F_{p}(\Delta x_1) = \int d \Delta x_{2} F(\Delta x_1, \Delta x_2)$.}
This function
is the same for the two paths of the PDC and it is well approximated by a Gaussian profile. 
It gives the probability
to detect a photon vs the spatial coordinates, so its width is the PDC width. 
From a numerical approach we obtain $w_p \approx 20$ pixels.
This number is confirmed by a direct measure of the PDC profile. 
We note also that this width is directly connected with the Sinc function and is only weakly
dependent on the collection spatial efficiency. This is a consequence of the 
fact that in our experimental scheme the PDC cone is forced to remain in a little area
due to the presence of the lens $f$.
\\ \\
Now we can face the derivation of the correlation length $w_{\text{\tiny cp}}$. This quantity is of fundamental importance in our work and it gives the probability to detect a photon within a definite interval when the other photon is found in a definite position (a pixel in our case). 
In particular, it is clear that in order to 
  define a proper $\varphi$ function, we need an experimental apparatus able to generate a correlation length of only few pixels and for this reason we use the configuration
with the lenses $f$ between the crystals and the SLM.
The correlation length \textcolor{black}{$w_{\text{\tiny cp}}$} has two contributions, one connected directly to the function $F$ and the other one connected with the pump dimension. The first contribution is the width 
\textcolor{black}{$\tilde{w}_{\text{\tiny pc}}$} of the function $\textcolor{black}{F_{cp}}=F(\Delta x_1, 0)$. About this function it is important to say that this width
doesn't change if we integrate the position 2 along the dimension of one pixel. $\tilde{w}_{\text{\tiny cp}}$ increases with the spectrum width and the profile of $F_{cp}$
is well reproduced by a Gaussian when the spectrum width is smaller 
than $15$ nm and it is well reproduced by a \textcolor{black}{super-Gaussian} with $n=4$
for bigger spectrum width. This result depends by the fact that with our spectrum selector we obtain a quasi rectangular profile of the PDC spectrum.}
\textcolor{black}{The second contribution is related with the pump spot dimension. 
About this we have to consider that the PDC is generated not only in one point in 
the transversal direction but along the pump profile \cite{jha10}. The point  is that the spatial coherence properties of the pump are directly transferred into the PDC.
In a naive picture we can say that the single mode of the pump is transferred into the single mode (defined by the direction $\theta$) of the PDC.
This means that the lens $f$ \textcolor{black}{focuses} this single mode on the SLM plane with a dimension $w^0_{cp}=\frac{\lambda_0 f}{\pi w_{\text{\tiny pump}}} \approx 1$ pixel 
where $\lambda_0=810$ nm. Without \textcolor{black}{focusing},
 $w^0_{\text{\tiny cp}} $ would be equal to the pump dimension, indeed in our case we have a well collimated pump.
An alternative scheme would be to use a \textcolor{black}{focused} pump but we note that in this case we have to put the lens before the crystals obtaining in turn
a bigger dimension on the SLM plane.
Finally, considering these two contributions, we have $w_{\text{\tiny cp}}=\sqrt{\left(\tilde{w}_{\text{\tiny cp}}\right)^2+\left( w^0_{\text{\tiny cp}} \right)^2}$. \textcolor{black}{So we can write $F(\Delta x_1, \Delta x_2)=e^{-\frac{2(\Delta x_1 - \Delta x_2)^n}{w^n_{\text{\tiny cp}}}} e^{-\frac{2(\Delta x_1)^2}{w_p^2}}e^{-\frac{2(\Delta x_2)^2}{w^2_p}}$ (without considering a normalization factor). In order to obtain the Equation 5 of the main text,
we have only to demonstrate that $|f_{jk}|^2 = F(\Delta x_1, \Delta x_2)$. And we can easily to see that this equality is assured by the fact that phase $\Phi$ is not a function of $\omega$.}
}
\\
\\
{\bf Measure of the coherence factor} $-$
\label{app2}
If the system is in the state $\rho_{\text{\tiny S}}(t)$, and the polarizers are both at $45^\circ$, the detection probability \textcolor{black}{(considering \textcolor{black}{the quantum efficiency} QE=1) is:
\begin{equation}
p_{++}=\frac{1}{4}\left(1+p\Re{\Gamma}\right),\nonumber
\end{equation}
while, if one polarizer is at $45^\circ$ and the other at $135^\circ$, it results$
p_{+-}=\frac{1}{4}\left(1-p\Re{\Gamma}\right)$.
Then, the coincidence counts for second in the two cases are:
\begin{equation}
\begin{dcases}
N_{++}=N_0 \left( 1+p \left( \Re{\Gamma} \right) \right)\\
N_{+-}=N_0 \left( 1-p \left( \Re{\Gamma} \right) \right)
\end{dcases},\nonumber
\end{equation}
where $N_0$ is obtained directly from the experimental counts and 
 it takes into account the spatial-spectral quantum
efficiencies of the detection system. So we can infer  information about $\Gamma$ from the visibility:
\begin{equation}
V=\abs{\frac{N_{++}-N_{+-}}{N_{++}+N_{+-}}}=p\abs{\Re{\Gamma}}\nonumber
\end{equation}}
\textcolor{black}{In the ideal case (without undersampling effect \cite{rossi17})  $\Gamma$ is a real quantity.
 This case can be experimentally recovered taking the array of noise phases $\varphi$ with zero mean. In the graphs in Figures 3 and 4 (in the main text) we show the comparison between the theoretical curves of $\abs{\Re{\Gamma}}$ and  $\abs{\Gamma}$ to put in evidence the effectiveness of this method. }
\\
\\
{\bf Measure of the number of correlated pixels} $-$
\label{app3}
Let's introduce in the first half of the SLM a rectangular function which switches $\pm\pi/4$ every $n_r=5$ pixels, and in the second half the same function shifted by $h$. Therefore $\Gamma$ is a function of $h$. By the measurements of $N_{++}=N_{++}(h)$ and $N_{+-}=N_{+-}(h)$ for $h=-10,-9,...,9$, where each point is an average of $4$ measures and each measure has an acquisition time of $8$ s, we calculate $V=V(h)$, as shown 
in the left panel of Fig. \ref{coherencelength}. The visibility of $V(h)$ is a decreasing function of the number of correlated pixels $w_{\text{\tiny cp}}$ and it can be simulated as shown in the central panel of Fig. \ref{coherencelength}, so that from the experimental value of $\text{Vis}(V(h))$ one can extrapolate $w_{\text{\tiny cp}}$. Repeating the experiment for different apertures of the spectral selector, we obtain the number of correlated pixels as a function of $\Delta\lambda$ (see the right panel 
of Fig. \ref{coherencelength}). 
The experimental data undergo a saturation for small $\Delta\lambda$ because of the effect of the transversal width of the pump.
\begin{figure}[!h!]
	\centering
	\includegraphics[width=0.3\textwidth]{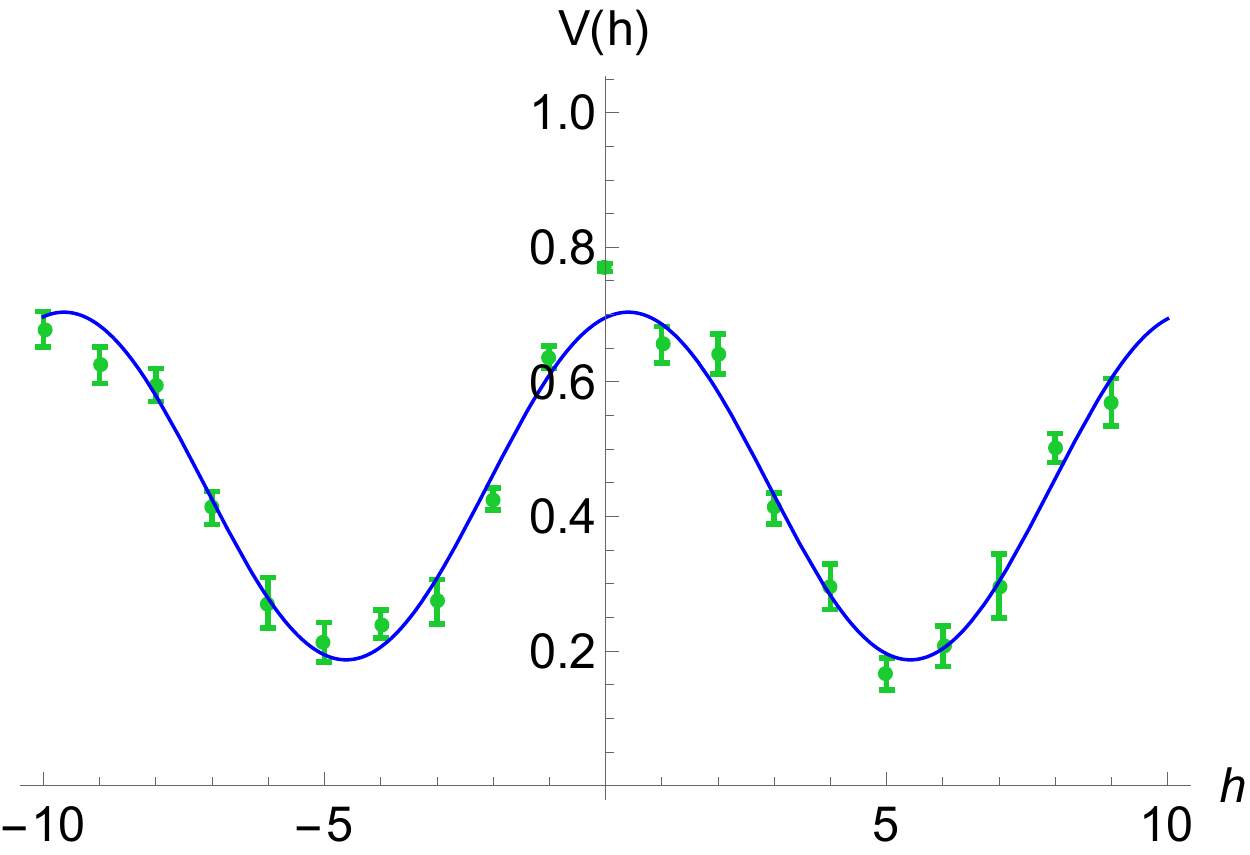}
	\includegraphics[width=0.3\textwidth]{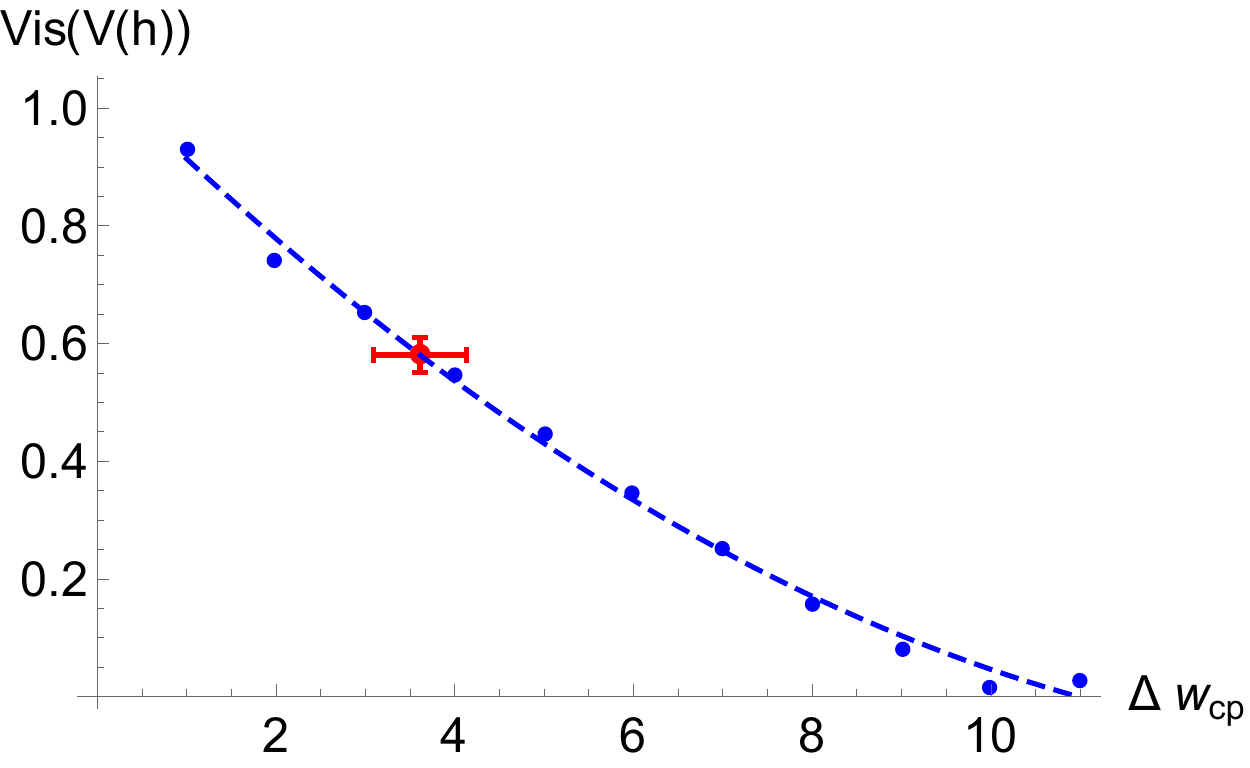}
	\includegraphics[width=0.3\textwidth]{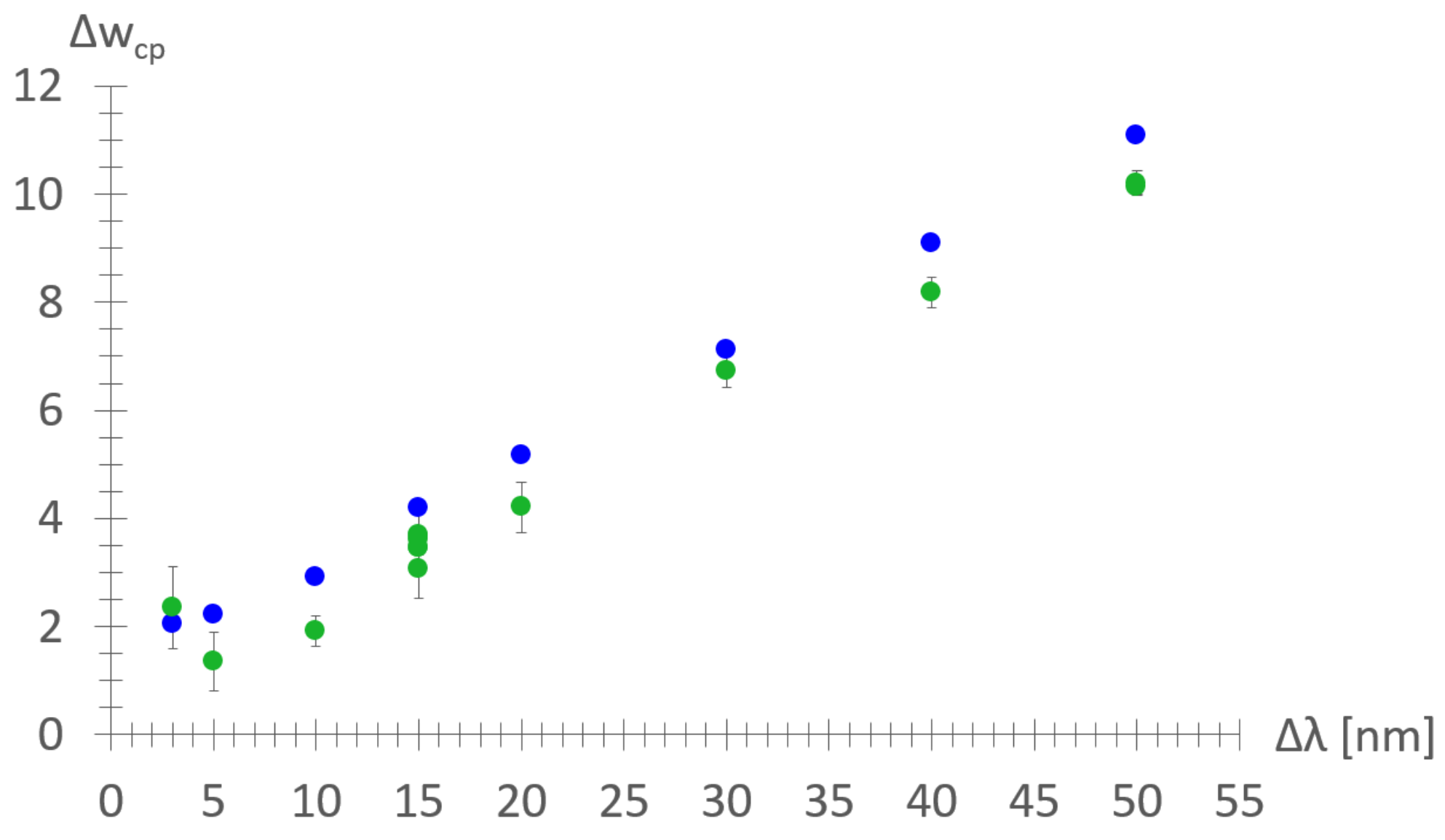}
	\caption{Left panel: The green dots represent the measured values of $V(h)$ for $\Delta\lambda=15$ nm and $p=0.927$, while the blue line is the fitting sine wave. In the present case, it results $\text{Vis}(V(h))=0.58\pm0.03$.
Center panel: The blue dots represent the value of $\text{Vis}(V(h))$ obtained by simulation for $\Delta\lambda=15$ nm and $q=0.927$, while the blue dashed line is the fitting polynomial curve. The red dot is the measured value of visibility. In the present case, the extrapolated number of correlated pixel is $\textcolor{black}{w_{\text{\tiny cp}}=3.1}\pm0.5$. Right Panel: Number of correlated pixels as a function of the spectral width of the PDC: the experimental data in green, the simulated values in blue.	
	}
	\label{coherencelength}
\end{figure}
\end{widetext}

\end{document}